\documentclass[%
 aps,
 prl,
nofootinbib,
nobibnotes,
amsmath,amssymb,
superscriptaddress,
 reprint,%
floatfix]{revtex4-2}

\usepackage{graphicx}
\usepackage{appendix}

\usepackage{dcolumn}
\usepackage{bm}
\usepackage[utf8]{inputenc}
\usepackage[T1]{fontenc}
\usepackage{mathptmx}
\usepackage{etoolbox}
\usepackage{hyperref}

\usepackage{xcolor}
\definecolor{MainColor}{HTML}{686CB2}

\hypersetup{
	colorlinks,
 linkcolor=MainColor,
 citecolor=MainColor,
 urlcolor=MainColor
}

\usepackage[version=3]{mhchem}

\usepackage{array}

\usepackage{comment}
\usepackage{cleveref}

\crefrangeformat{equation}{Eqs.~(#3#1#4) --~(#5#2#6)}

\newcommand{\dd}{\mathrm{d}}

\usepackage[colorinlistoftodos]{todonotes}
\usepackage[labelformat=simple,caption=false]{subfig}

\makeatletter
\newlength{\sfp@hseplen}\newlength{\sfp@vseplen}
\define@cmdkey{subfigpos}[sfp@]{pos}[ul]{}
\define@cmdkey{subfigpos}[sfp@]{font}[\small]{}
\define@cmdkey{subfigpos}[sfp@]{vsep}[2\baselineskip]{\setlength{\sfp@vseplen}{\sfp@vsep}}
\define@cmdkey{subfigpos}[sfp@]{hsep}[10pt]{\setlength{\sfp@hseplen}{\sfp@hsep}}
\newcommand{\subfigimg}[3][,]{%
  \setkeys{Gin,subfigpos}{pos,font,vsep,hsep,#1}
  \setbox1=\hbox{\includegraphics{#3}}
  \ifnum\pdfstrcmp{\sfp@pos}{ul}=0
    \leavevmode\rlap{\usebox1}
    \rlap{\hspace*{\sfp@hsep}\raisebox{\dimexpr\ht1-\sfp@vsep}{\sfp@font{#2}}}
    \phantom{\usebox1}
  \else\ifnum\pdfstrcmp{\sfp@pos}{ur}=0
    \leavevmode\usebox1
    \llap{\raisebox{\dimexpr\ht1-\sfp@vsep}{\sfp@font{#2}}\hspace*{\sfp@hsep}}
  \else\ifnum\pdfstrcmp{\sfp@pos}{lr}=0
    \leavevmode\usebox1
    \llap{\raisebox{\sfp@vsep}{\sfp@font{#2}}\hspace*{\sfp@hsep}}
  \else
    \leavevmode\rlap{\usebox1}
    \rlap{\hspace*{\sfp@hseplen}\raisebox{\sfp@vsep}{\sfp@font{#2}}}
    \phantom{\usebox1}
  \fi\fi\fi
}

\makeatletter
\def\@email#1#2{%
 \endgroup
 \patchcmd{\titleblock@produce}
  {\frontmatter@RRAPformat}
  {\frontmatter@RRAPformat{\produce@RRAP{*#1\href{mailto:#2}{#2}}}\frontmatter@RRAPformat}
  {}{}
}%
\makeatother
\begin{document}

\preprint{AIP/123-QED}

\title[]{Pressure-Gated Microfluidic Memristor for Pulsatile Information Processing}

%
\title[]{Pressure-Gated Microfluidic Memristor for Pulsatile Information Processing}\author{A. Barnaveli}
\thanks{Author to whom correspondence may be addressed: a.barnaveli@uu.nl}
\affiliation{Institute for Theoretical Physics, Utrecht University,  Princetonplein 5, 3584 CC Utrecht, The Netherlands}
\author{T. M. Kamsma}
\affiliation{Institute for Theoretical Physics, Utrecht University,  Princetonplein 5, 3584 CC Utrecht, The Netherlands}
\affiliation{Mathematical Institute, Utrecht University, Budapestlaan 6, 3584 CD Utrecht, The Netherlands}
\author{W. Q. Boon}
\affiliation{Institute for Theoretical Physics, Utrecht University,  Princetonplein 5, 3584 CC Utrecht, The Netherlands}
\author{R. van Roij}%
\affiliation{Institute for Theoretical Physics, Utrecht University,  Princetonplein 5, 3584 CC Utrecht, The Netherlands}

\date{\today}

\begin{abstract}
A hitherto unexploited characteristic feature of emerging iontronic devices for information processing is the intrinsic mobility of the medium (water) of dissolved ions in aqueous electrolytes, which therefore not only respond to voltage but also to pressure. Here we study a microfluidic memristor, in the form of a conical channel, exposed to simultaneously applied time-dependent voltage and pressure drops, through numerical solutions of the Poisson-Nernst-Planck-Stokes equations for ion and fluid transport. We show that the channel's memristive properties can be enhanced, reduced or instantaneously reset by a suitable pressure, and we leverage this finding with two examples of time series processing of simultaneously applied voltage and  pressure pulses. We not only show that the distinction between different voltage time series can be improved by enhancing the conductance response with corresponding pressure pulses, but also that the bandwidth of information transfer through the channel can be doubled by letting the pressure pulses represent a second independent time series. 

\end{abstract}

\maketitle

Memristors (memory-resistors) exhibit a conductance that depends on past current or voltage inputs \cite{strukov2008missing,chua1971memristor}. They have drawn significant interest, predominantly driven by their ability to emulate neuronal processes in neuromorphic (brain-inspired) circuits \cite{zhu2020comprehensive,sangwan2020neuromorphic,schuman2017survey}.  This has led to the employment of memristive devices for information processing \cite{schuman2022opportunities}, for instance by emulating weights in neural networks \cite{xia2019memristive}, consequently circumventing the von Neumann bottleneck by co-locating memory and processing as in the brain \cite{ielmini2018memory}. These forms of \textit{neuromorphic computing} are considered to be promising candidates for addressing the unsustainably rising energy consumption and the unmanageably large amounts of generated data \cite{mehonic2022brain,schuman2022opportunities}. For these reasons a diversity of solid-state memristors has by now been explored as possible promising circuit components for neuromorphic and analog computations \cite{jo2010nanoscale,yang2013memristive,strukov2008missing,kuzum2012nanoelectronic,prezioso2015training,lanza2022memristive,wang2018fully,schuman2017survey,sangwan2020neuromorphic,zhu2020comprehensive}. 

However, these solid-state memristors for neuromorphic circuits also reveal some fundamental disparities with the brain, which employs ions and molecules in an aqueous environment rather than electrons and holes in a solid semiconductor, thereby utilising chemical regulation as well as multiple information carriers in parallel. These fundamental differences result in tangible challenges, e.g.\ the fast dynamics of solid-state may cause problems in processing time series with comparatively slower time scales such as biological signals \cite{Chicca2020ASystems}. In order to address this disparity, devices have been proposed that feature electrochemical coupling between (fast) electrons in metallic/semiconducting materials and (slow) dissolved ions/protons \cite{zhu2020comprehensive,lanza2022memristive,waser2009redox,waser2007nanoionics,sawa2008resistive,kim2011nanofilamentary,valov2011electrochemical,Wang2022DynamicBehaviour,VanDeBurgt2018OrganicComputing,Harikesh2022OrganicSpiking,Harikesh2023IonTunable,Luo2023HighlyDynamics}. Interestingly, an emerging class of memristors operates solely on the basis of aqueous electrolytes without any electronic component \cite{powell2011electric,zhou2023nanofluidic,wang2012transmembrane,li2015history,wang2014physical,wang2016dynamics,wang2017correlation,sheng2017transporting,brown2020deconvolution,brown2022selective,brown2021higher,wang2018hysteresis,ramirez2021negative,martinez2023fluidic,robin2023long,xiong2023neuromorphic,ramirez2024neuromorphic,ratschow2022resonant}. Various functionalities have recently been extracted from these fluidic iontronic devices, for instance features of synaptic plasticity \cite{ramirez2023synaptical,han2023iontronic}, chemical regulation \cite{xiong2023neuromorphic,robin2023long,wang2024aqueous}, theoretical proposals of neuron-like spiking \cite{robin2021principles,kamsma2023iontronic,kamsma2024advanced}, implementations for traditional truth tables \cite{emmerich2023ionic,sabbagh2023designing,li2023high}, and initial demonstrations of neuromorphic computing \cite{kamsma2023brain}. While these results are promising, the advancement of aqueous neuromorphic devices is still in the early stages and more work is required to convert the unique features of iontronics to tangible benefits \cite{han2022iontronics,xie2022perspective,noy2023fluid}.

\begin{figure} 
	\centering
	\includegraphics[width=0.99\linewidth]{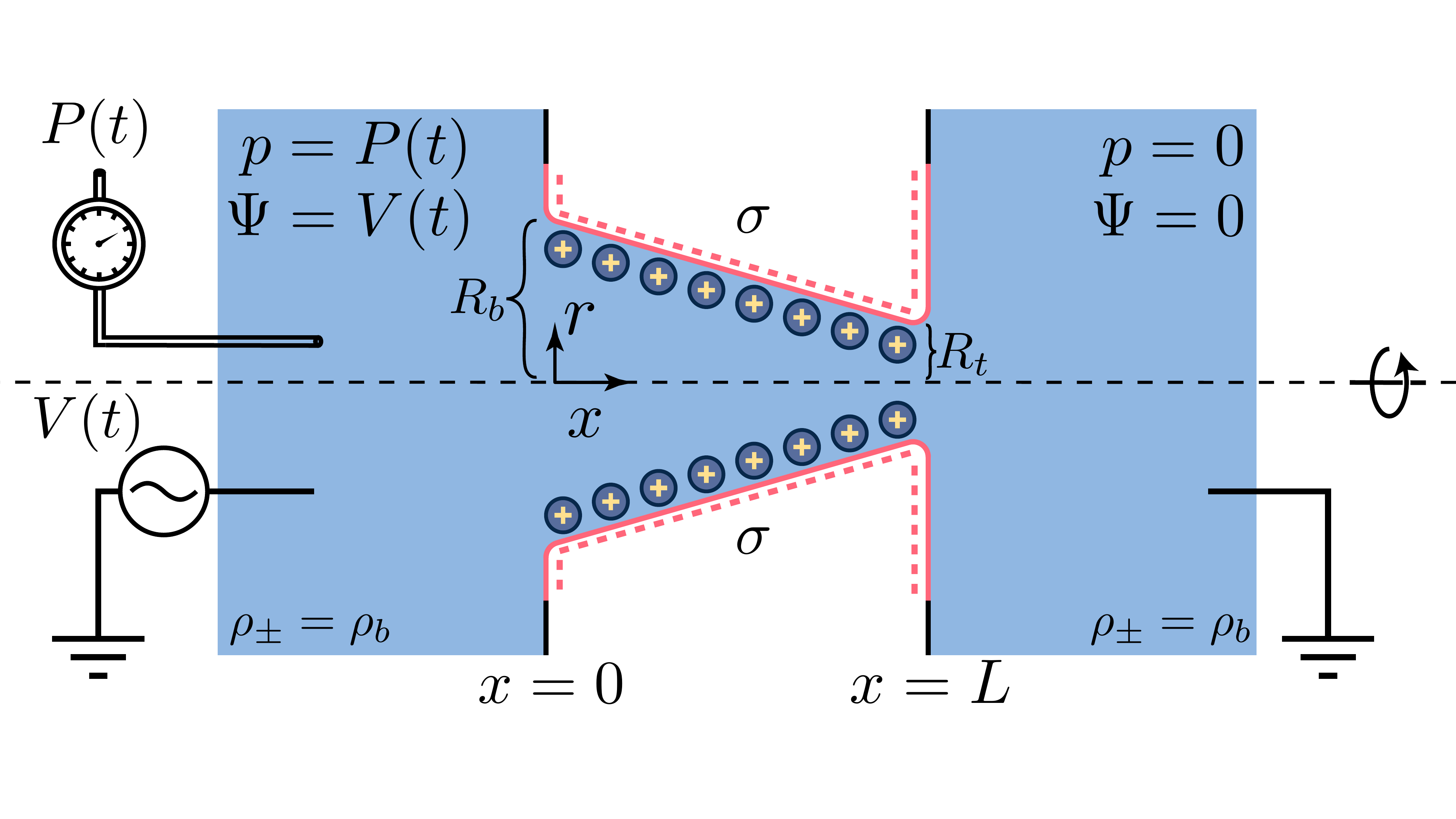}
	\caption{Schematic (not to scale) of an azimuthally symmetric conical channel of length $L$, base radius $R_{\mathrm{b}}$, and tip radius $R_{\mathrm{t}}<R_{\mathrm{b}}$, connecting two bulk reservoirs of an aqueous 1:1 electrolyte with equal ionic bulk concentration $2\rho_{\mathrm{b}}$. The channel walls carry a negative surface charge density $\sigma$. A time-dependent potential drop $V(t)$ and pressure drop $P(t)$ are simultaneously applied across the channel, inducing an electrolyte volume flow $Q(t)$ and an ionic charge current $I(t)$ that we calculate on the basis of Poisson-Nernst-Planck-Stokes equations, yielding a pressure-tunable time-dependent electric conductance $g(t)=I(t)/V(t)$, that can be used to transfer information. 
 } 
 \label{Fig:Cone}
\end{figure}

In this Letter we study iontronic memristors in the form of microfluidic conical channels and leverage the unique property of iontronics that the charge-carrying medium (water) is mobile itself. In response to an applied DC voltage drop, it is well established experimentally and theoretically that conical channels with charged channel walls (i) exhibit ionic current rectification, i.e.\ the channel conductance depends on the magnitude and the polarity of the applied (static) voltage \cite{white2008ion,wen2019rectification,proctor2021theory,woermann2003electrochemical,woermann2004electrochemical,kovarik2009effect,lin2018voltage}, and (ii) act as a pressure-gated transistor because a simultaneously applied pressure drop can strongly affect the electric current \cite{lan2011pressure,poggioli2019beyond,jubin2018dramatic,boon2022pressure}. These effects were both quantitatively explained by a theoretical model that describes how the conductance stems from a voltage-dependent steady-state salt concentration profile in the conical channel \cite{boon2022pressure}. This concentration polarisation not only depends on the polarity of the applied voltage but also on the net fluid flow with pressure-induced (Poiseuille-like) and voltage-induced (electro-osmotic) contributions \cite{boon2022pressure}. 
Recently also the timescale for the voltage-induced build up of concentration polarisation was identified \cite{kamsma2023iontronic}, which provided a quantitative explanation for experimentally found memristive effects in conical channels \cite{wang2012transmembrane,li2015history,wang2014physical,wang2016dynamics,wang2017correlation,sheng2017transporting,brown2020deconvolution,brown2022selective,brown2021higher,wang2018hysteresis,ramirez2023synaptical,ramirez2024neuromorphic}. In this Letter we combine these theoretical insights \cite{boon2022pressure,kamsma2023iontronic} to show that \textit{pulsatile} pressure drops can be combined with AC voltages to not only amplify and reset the memristive properties of a microfluidic conical channel, but also to increase its signalling bandwidth. Although pressure sensors have been integrated into a solid-state device to induce resistive switching \cite{keshari2024light}, pressure sensitivity is not an intrinsic feature of solid-state devices, which usually only allow for pressure signaling via connections to dedicated sensors \cite{zhu2022artificial,zhang2020artificial,liu2023near}. In contrast, fluid flow, and hence pressure sensitivity, is an inherent property of fluidic iontronic memristors, therefore the features we present here represent results that we believe will be of relevance across the various emerging iontronic devices.

The system of interest, illustrated in Fig.~\ref{Fig:Cone}, with experimentally realistic parameters, comprises a long azimuthally symmetric conical channel, with axial and radial coordinates $x$ and $r$, of length $L=9.8\text{\,$\mu$m}$, a base radius $R_b=446\text{\,nm}$ in the plane $x=0$, and a tip radius $R_t=98\text{\,nm}$ in the plane $x=L$. Its cross section radius reads $R(x)=R_b -(x/L)(R_b-R_t)$ for $x\in[0,L]$. The channel connects two incompressible aqueous 1:1 electrolyte reservoirs at $x<0$ and $x>L$, each at equal ionic bulk concentration $2\rho_b$ at room temperature $T=293\text{\,K}$, viscosity $\eta=1\text{\,mPas}$, mass density $\rho_m=1\text{\,kg/L}$, dielectric constant $\epsilon=80$, default bulk salt concentration $\rho_b=1\text{\,mM}$ and ionic diffusion coefficient $D=1\text{\,$\mu$m}^2/\text{ms}$. The walls of the channel carry a homogeneous surface charge density $\sigma$, such that the point-like and mobile cations and anions form an electric double layer in the channel. The channel walls are blocking and we exclude Faradaic processes. The surface charge is set to $\sigma\approx -3.4\text{\,mC/m$^2$}$, which yields an equilibrium zeta potential of $40\text{\,mV}$ to mimic a silica surface in contact with an aqueous $1:1$ electrolyte with a Debye screening length $\lambda_D\simeq 9.8\text{\,nm}$ \cite{iler1955colloid}. Note that the channel dimensions, which are in the regime of weak double-layer overlap since $\lambda_D\ll R_t$, satisfy $R_t/R_b=0.22$, found to be optimal for current rectification \cite{boon2022pressure}. On the far side of both reservoirs we impose fixed ionic bulk concentrations of $\rho_{\pm}=\rho_{\mathrm{b}}=1\text{\,mM}$. We impose a time-dependent voltage $V(t)$ and pressure $P(t)$ at the far side of the reservoir connected to the base, while the far side of the tip reservoir is grounded and at (arbitrary) reference pressure zero. 

The above-mentioned channel features lead to an electric potential profile $\Psi(x,r,t)$, a pressure profile $p(x,r,t)$, a fluid flow with velocity field $\mathbf{u}(x,r,t)$, and ionic fluxes $\mathbf{J}_{\pm}(x,r,t)$. To resolve the fluid and ionic dynamics in the channel we solve the full set of Poisson-Nernst-Planck-Stokes (PNPS) equations for diffusive, conductive, and advective transport
\begin{gather}
	\nabla^2\Psi=-\frac{e}{\epsilon_0\epsilon}(\rho_+-\rho_-\big),\label{eq:Poisson}\\
	\dfrac{\partial\rho_{\pm}}{\partial t}+\nabla\cdot\mathbf{J}_{\pm}=0,\label{eq:Continuity}\\
 \mathbf{J}_{\pm}=-D_{\pm}\left(\nabla\rho_{\pm}\pm\rho_{\pm}\frac{e\nabla \Psi}{k_{\mathrm{B}}T}\right)+\mathbf{u}\rho_{\pm},\label{eq:NernstPlanck}\\
	\rho_m\dfrac{\partial\mathbf{u}}{\partial t}=\eta\nabla^2\mathbf{u}-\nabla p-e(\rho_+-\rho_-)\nabla \Psi;\qquad\nabla\cdot\mathbf{u}=0.\label{eq:Stokes}
\end{gather}
The Poisson Eq.~(\ref{eq:Poisson}) accounts for electrostatics. The conservation of ions is ensured through the continuity Eq.~(\ref{eq:Continuity}), while the ionic fluxes are assumed to be described by the Nernst-Planck Eq.~(\ref{eq:NernstPlanck}), which combines Fickian diffusion, Ohmic conduction, and Stokesian advection. Finally the force balance on the (incompressible) fluid is described by the Stokes Eq.~(\ref{eq:Stokes}). To close the system of Eqs~(\ref{eq:Poisson})-({\ref{eq:Stokes}}) we impose no-slip, no-flux, and Gauss' law boundary conditions on the walls of the system, i.e.\ $\mathbf{u}=0$, $\mathbf{n}\cdot\mathbf{J}_{\pm}=0$, and $\mathbf{n}\cdot\nabla\Psi=-\sigma/\epsilon_0\epsilon$, respectively, with $\mathbf{n}$ the wall's inward normal vector.

Of particular interest here is the time-dependent electric current $I(t)=2\pi e\int_0^{R(x)} \big({\bf J}_+(x,r,t) - {\bf J_-}(x,r,t) \big)\cdot\hat{\bf{x}} r \dd r$, with $\hat{\bf{x}}$ the unit vector in the positive $x$-direction, and the volume fluid flow $Q(t)$, driven through the channel by a simultaneously applied time-dependent potential $V(t)$ and pressure $P(t)$. Numerically we find that the capacitance of the channel is so small that the $x$-dependence of $I(t)$ is negligible (see SM). The charge flux ${\bf J}_+ - {\bf J_-}$, from which the current follows, depends on the channel salt concentration $\rho_+ +\rho_-$, which can be (dynamically) increased or decreased by (simultaneously) applied potential and pressure drops \cite{boon2022pressure,kamsma2023iontronic}. For a purely voltage driven process, the typical salt concentration polarisation timescale was found to be well-estimated by \cite{kamsma2023iontronic,kamsma2023unveiling}
\begin{equation} \label{eq:Timescale_Diffusive}
    \tau=\frac{L^2}{12D},
\end{equation}
which results in a voltage-dependent conductance memory retained over a time $\sim\tau$. Here $\tau=8.33\text{\,ms}$ for the present parameters. However, steady-state salt concentration polarisation also strongly depends on the fluid flow field \cite{boon2022pressure}, which develops much faster as we will see. Therefore we expect that the dynamic voltage-dependent conductance can be tuned by a quasi-instantaneous response of the fluid flow to an applied pressure drop $P(t)$. To investigate the resulting interesting interplay between voltage and pressure drop $V(t)$ and $P(t)$, we numerically solve the PNPS equations using finite-element methods of COMSOL\textregistered \:.

\begin{figure} [ht!]
\captionsetup[subfigure]{labelformat=empty}
\subfloat[\label{Fig:Hysteresis_Comparison}]{\subfigimg[width=\linewidth,pos=tl,vsep=1.05cm,hsep=1.75cm]{\large (a)}{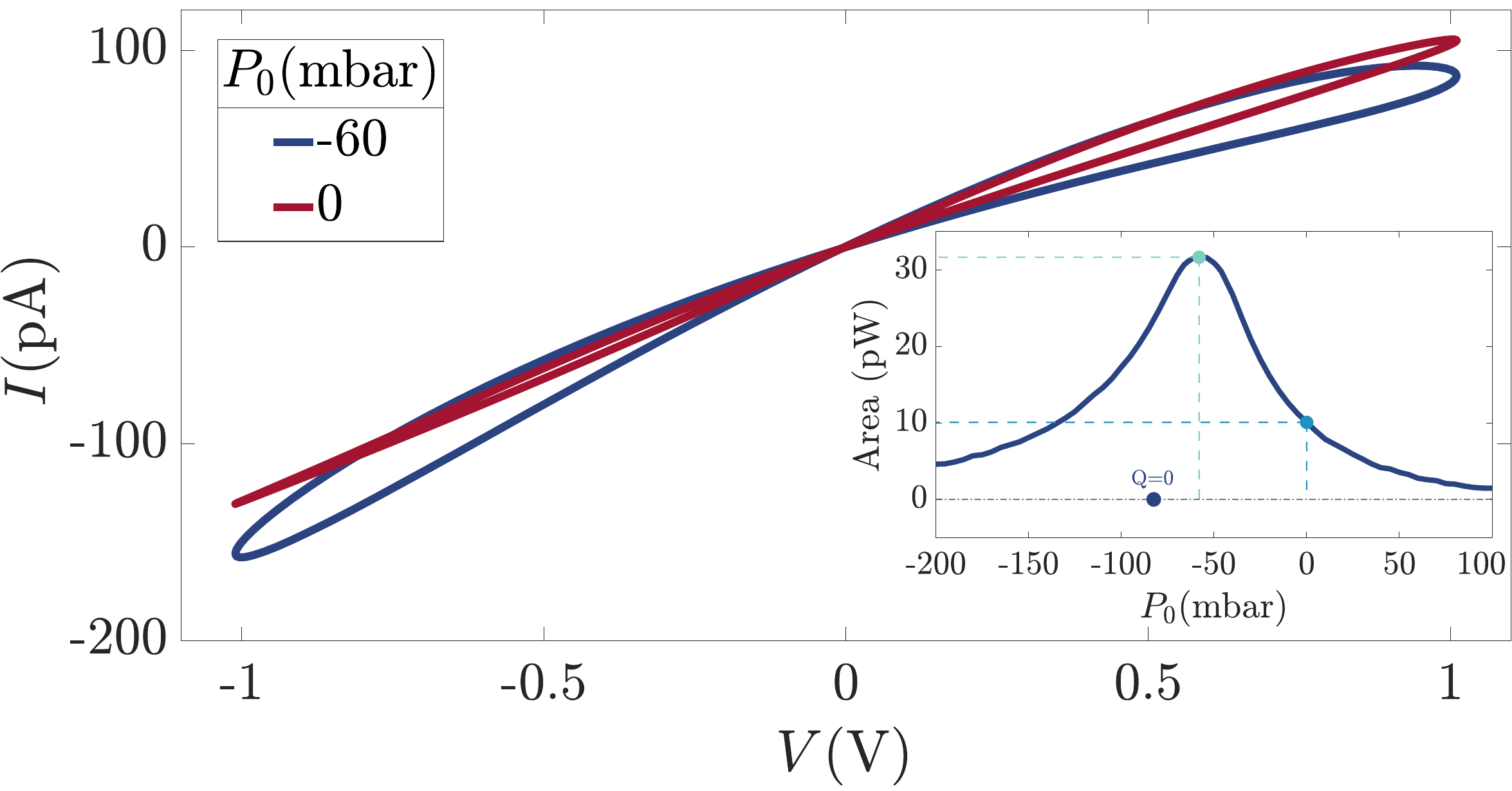}}\vspace{-1.37cm}\quad
\subfloat[\label{Fig:Memory_Reset}]{\subfigimg[width=\linewidth,pos=tl,vsep=1.35cm,hsep=1.25cm]{\large (b)}{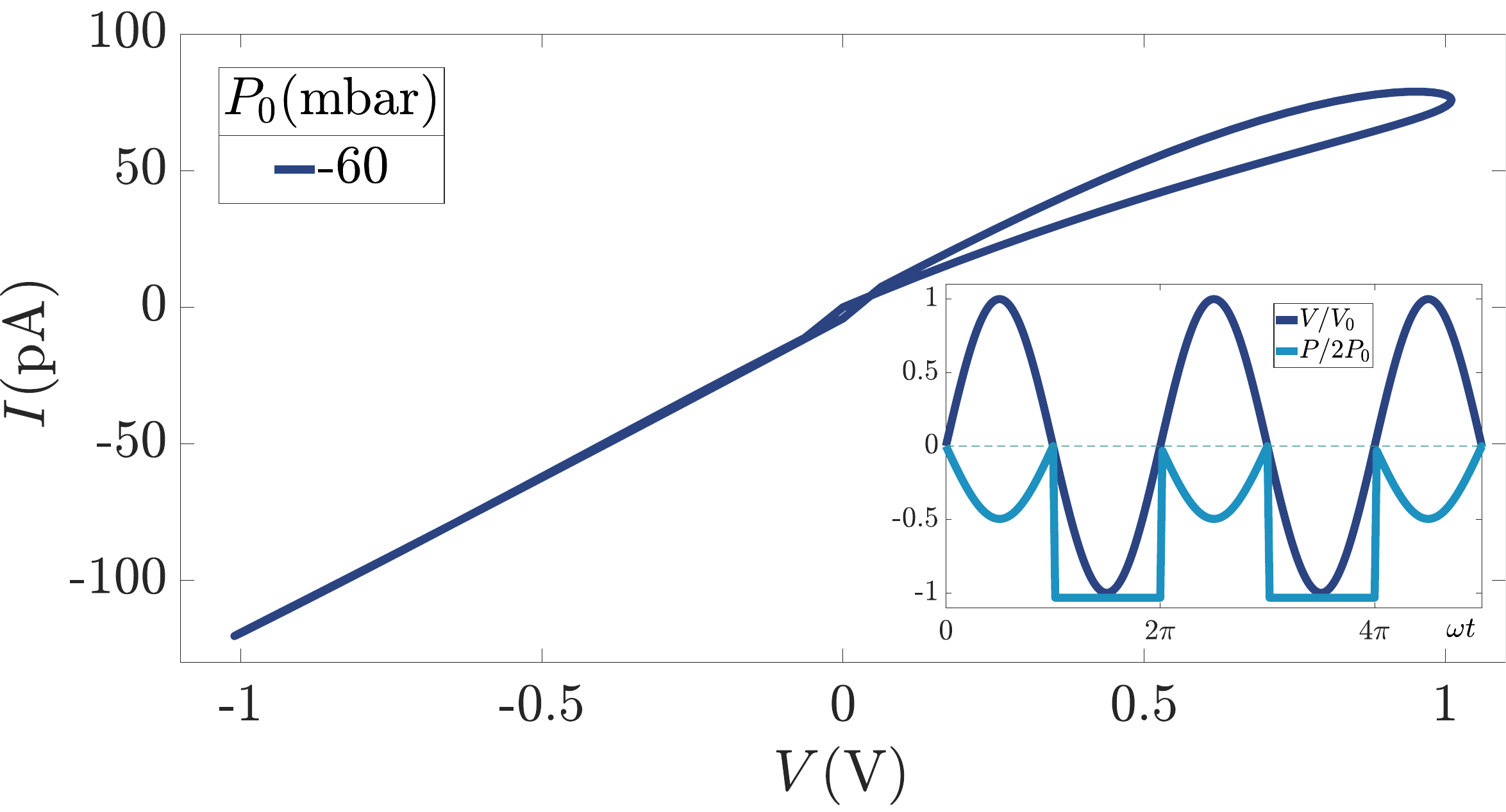}}
\caption{Current-voltage hysteresis loops of a conical channel, for our standard parameter set (see text), driven by a harmonic voltage drop (amplitude $V_0=1\text{ V}$ and frequency 25 Hz) combined with a simultaneously applied pressure-drop.\textbf{(a)} The pressure drop is also harmonic with amplitudes $P_0=-60\text{\,mbar}$ (blue) and $P_0=0$ (red), the inset showing the dependence of the area enclosed by the hysteresis loop on $P_0$. \textbf{(b)} The applied pressure $P(t)$, shown in the inset, equals $P_0\sin(\omega t)$ with $P_0=-60\text{ mbar}$ as in (a) if $V(t)>0$ and a constant $-120\text{ mbar}$ if $V(t)<0$, such that the memristive response at $V(t)>0$ is reset into an Ohmic response at $V(t)<0$.}
 \label{Fig:Voltammetry}
\end{figure}

We first study harmonic voltage- and pressure drops given by $V(t)=V_0\sin(\omega t)$ and $P(t)=P_0\sin(\omega t)$ at the judiciously chosen frequency of $\omega/2\pi =25\text{\,Hz}$, where $\omega\tau\simeq 1.3$ is close to unity and therefore close to the optimal frequency for observing memristive properties \cite{kamsma2023math}. We fix $V_0=1\text{\,V}$ and consider a variety of pressure amplitudes $P_0$.  
In Fig.~\ref{Fig:Hysteresis_Comparison} we plot the limit cycle of the parametric $I(t)$-$V(t)$ curve for $P_0=0$ (red) and $P_0=-60\text{\,mbar}$ (blue). For both pressures the current-voltage relation features a characteristic pinched hysteresis loop, with a closed loop at either voltage polarity and a crossing in the origin \cite{chua2014if}. However, the total area of the closed loop, a measure for the strength of the memristive effect, is seen to be much more pronounced for the non-zero pressure drop. The inset of Fig.~\ref{Fig:Hysteresis_Comparison} quantifies this by showing the $P_0$-dependence of the area of the hysteresis loop, featuring a well-defined peak at about $P_0=-60\text{\,mbar}$, where the area is enhanced by a factor larger than three compared to that at $P_0=0$; by contrast, the area is reduced compared to $P_0=0$ for $P_0>0$ and $P_0<-140\text{\,mbar}$, where a strong fluid flow washes out salt concentration polarisation and hence any significant conductance memory \cite{jubin2018dramatic,boon2022pressure}. Note that the zero-flow condition, $Q=0$ as indicated in the inset at $P_0\simeq -82\text{\,mbar}$, is close to the maximum, however not spot on. Nevertheless, the basic intuition that a static pressure washes out concentration polarisation or cancels the electro-osmotic fluid flow, thereby enhancing concentration polarisation, as detailed in Ref.~\cite{boon2022pressure}, largely carries over to the time-dependent regime. This is because the typical timescale for developing a Poiseuille-like velocity profile in a cylinder of radius $R$, as obtained from Stokes' Eq.~(\ref{eq:Stokes}), is given by $\tau'=R^2/(\eta/\rho_m)$, which equals $\tau^{\prime}=0.2\text{\,$\mu$s}$ for $R=R_b$ and is therefore many orders of magnitude smaller than $\tau$ and $1/\omega$. Therefore the fluid flow is essentially established instantaneously with pressure and acts very similar to a quasi-static flow \cite{womersley1955method}.    

Interestingly, besides allowing for manipulations of the memristance of a conical channel by applying pressure drops, the major difference between the characteristic time scales $\tau$ for ionic relaxation and $\tau'\ll\tau$ for momentum relaxation allows also for applications of pressure as a memory reset tool. 
This is illustrated in the current-voltage characteristic of Fig.~\ref{Fig:Memory_Reset} for the same system parameters and the same sinusoidal voltage $V(t)$ as in Fig.~\ref{Fig:Hysteresis_Comparison}, however with a sinusoidal pressure $P(t)=P_0\sin(\omega t)$ with $P_0=-60\text{ mbar}$ only if $V(t)>0$, and a constant pressure $P(t)=-120\text{ mbar}$ if $V(t)<0$, as plotted in the inset. This protocol for $P(t)$ reduces the flow and thus enhances the conductance memory (as shown in Fig.~\ref{Fig:Hysteresis_Comparison}) only for $V(t)>0$, while it increases the flow and washes out the concentration polarisation for $V(t)<0$, when it turns the channel essentially into an Ohmic conductor without conductance memory. Therefore, the pressure-induced flow for $V(t)<0$ acts as a memory eraser. The time-scale of the memory reset process is of order $\tau'$ and is therefore essentially instantaneous on the relevant memristive and electric time scale $\tau$.    

\begin{figure} [ht!]
\captionsetup[subfigure]{labelformat=empty}
\subfloat[\label{Fig:Conductance_Spread}]{\subfigimg[width=\linewidth,pos=tl,vsep=1.35cm,hsep=1.4cm]{\large(a)}{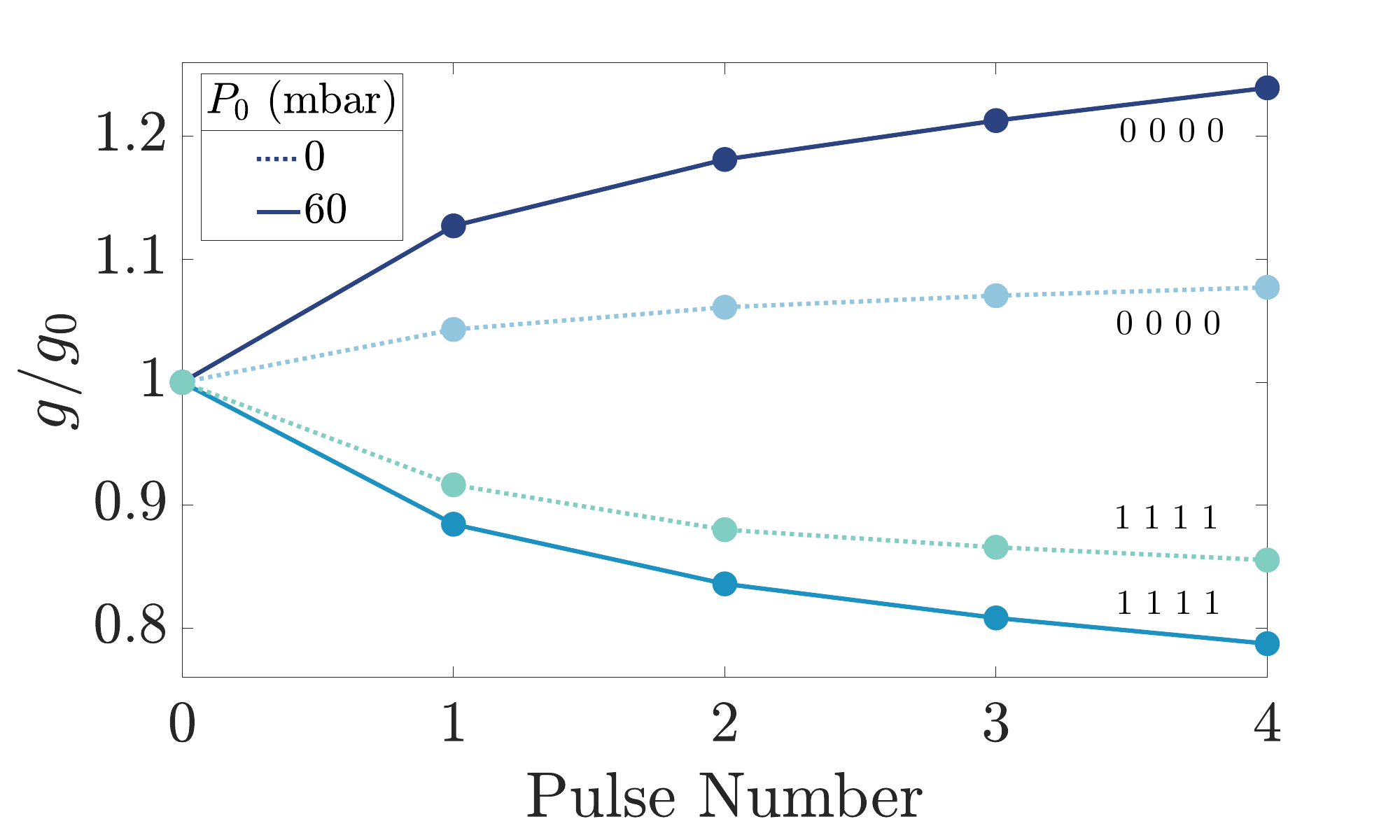}}\vspace{-1.3cm}\quad
\subfloat[\label{Fig:Conductance_4bits}]{\subfigimg[width=\linewidth,pos=tl,vsep=1.35cm,hsep=1.4cm]{\large(b)}{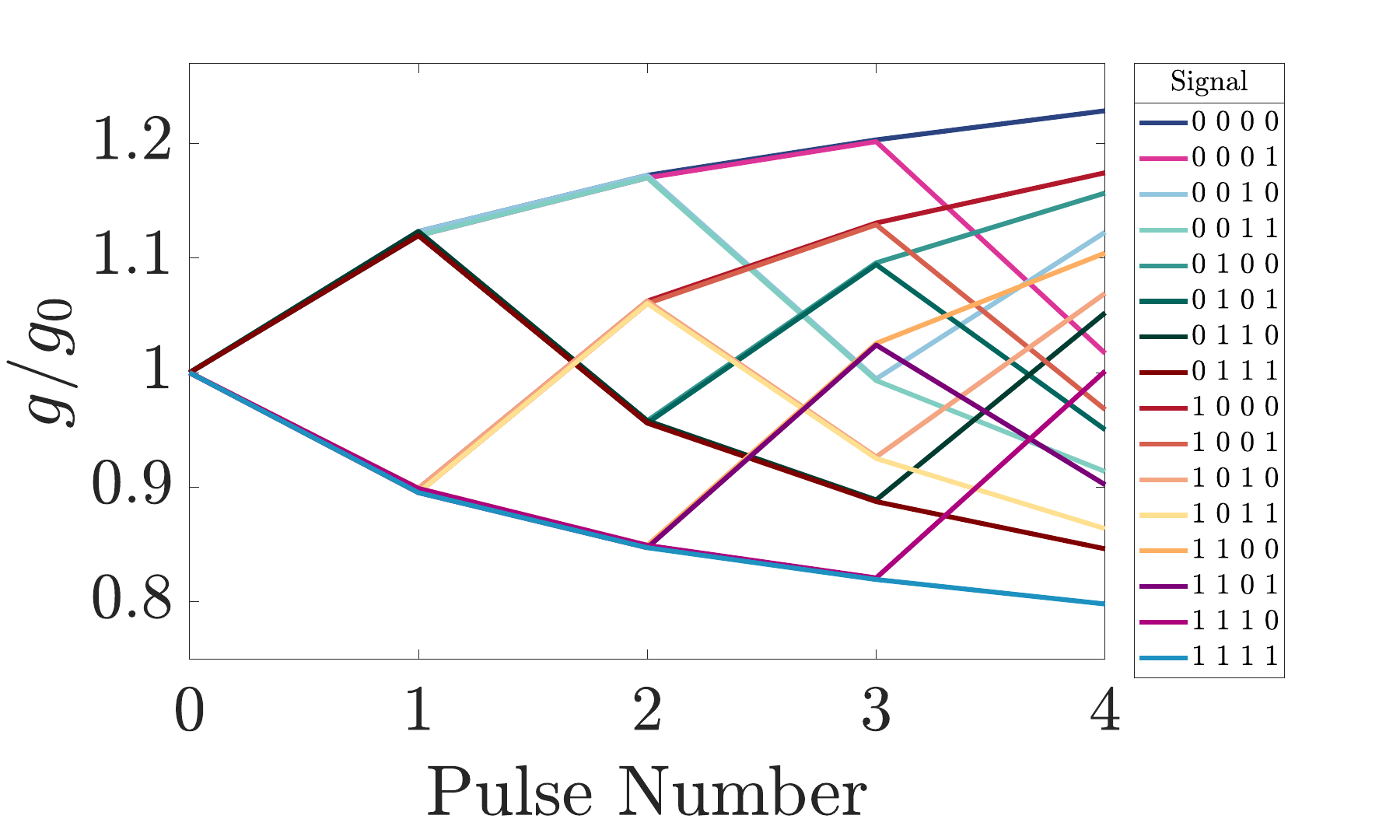}}
\caption{Relative channel conductance $g(t)/g(0)$ at $4t/\tau=0,1,2,3,4$ during a train (of duration $\tau$) of four voltage pulses, in (a) for the extremes of four
positive (1111) and four negative (0000) voltage pulses, without (dashed line) and with (solid line) a simultaneously applied train of four pressure pulses and in (b) for all 16 combinations of  four positive/negative voltage pulses with negative/positive simultaneous pressure pulses leading to a relatively wide range of 16 distinguishable conductances after the fourth pulse (see text). 
\label{Fig:Conductance}}
\end{figure}

Inspired by Refs.~\cite{du2017reservoir,kim2022implementation,pyo2022non,kamsma2023brain}, where memristors were shown to distinguish time series in the form of voltage pulses, we now study simultaneous voltage-pressure block pulses $\pm(V_0,P_0)$ with $V_0=1\text{\,V}$ and $P_0=-60\text{\,mbar}$, which we expect to optimise the electric response on the basis of our results of Fig.~\ref{Fig:Voltammetry}. We focus on trains of four of these pulses, all four of equal duration $\tau/8$ and all four preceded by a zero-voltage and zero-pressure interval of duration $\tau/8$, such that a train of four pulses together takes a total time $\tau$. By assigning a binary ``0'' and ``1'' to a negative and positive pulse, respectively, a train of four pulses can represent any of the 16 binary numbers between 0000 and 1111.
For each of the 16 possible trains, described by block signals $V(t)$ and $P(t)$, we solve the PNPS Eqs.~(\ref{eq:Poisson})-(\ref{eq:Stokes}) for our standard parameter set and calculate the electric conductance $g(t)=I(t)/V(t)$ of the channel at $4t/\tau=0, 1, 2, 3$, and $4$, so at the end of each pulse and at the start to find $g(0)$. In Fig.~\ref{Fig:Conductance_Spread} we plot the resulting normalised channel conductance $g(t)/g(0)$ for the two pulse trains representing 0000 and 1111, both at $P_0=-60\text{\,mbar}$ (solid lines) and $P_0=0$ (dashed). We observe, in agreement with our earlier findings, that negative/positive voltages increase/decrease the channel conductance, both for finite and zero pressure. However, the effect of the carefully chosen finite pressure $P_0=-60\text{\,mbar}$ considerably increases the range of achieved conductances (roughly twofold in this particular case). In Fig.~\ref{Fig:Conductance_4bits} we again show $g(t)/g(0)$ at the very end of each of the four pulses, but now only for $P_0=-60\text{\,mbar}$ and for all 16 possible trains. We see that each train, i.e.\ each bit string, is mapped onto a unique conductance $g(\tau)/g(0)$ that lies in between the two extreme ones of 0000 and 1111 (shown already in Fig.~\ref{Fig:Conductance_Spread}). The 16 unique conductances shown in Fig.~\ref{Fig:Conductance_4bits} are considerably closer to each other if only voltage pulses are used (see SM), therefore the pressure pulses significantly enhance the channel's capacity for distinguishing these time series.

Clearly, the pressure-induced widening of the conductance window enhances the separation of the different conductances $g(\tau)/g(0)$, which should facilitate their mutual distinction in conductance measurements of experimental realisations of these channels. In line with Ref.~\cite{kamsma2023brain}, where a colloid-filled tapered microchannel was employed as a synaptic device for reservoir computing with voltage-only pulses, an enhanced window of conductances and longer trains could be employed to increase the performance and computational capacity of the device or to decrease the number of required channels per computation.

\begin{figure} 
	\centering
	\includegraphics[width=0.99\linewidth]{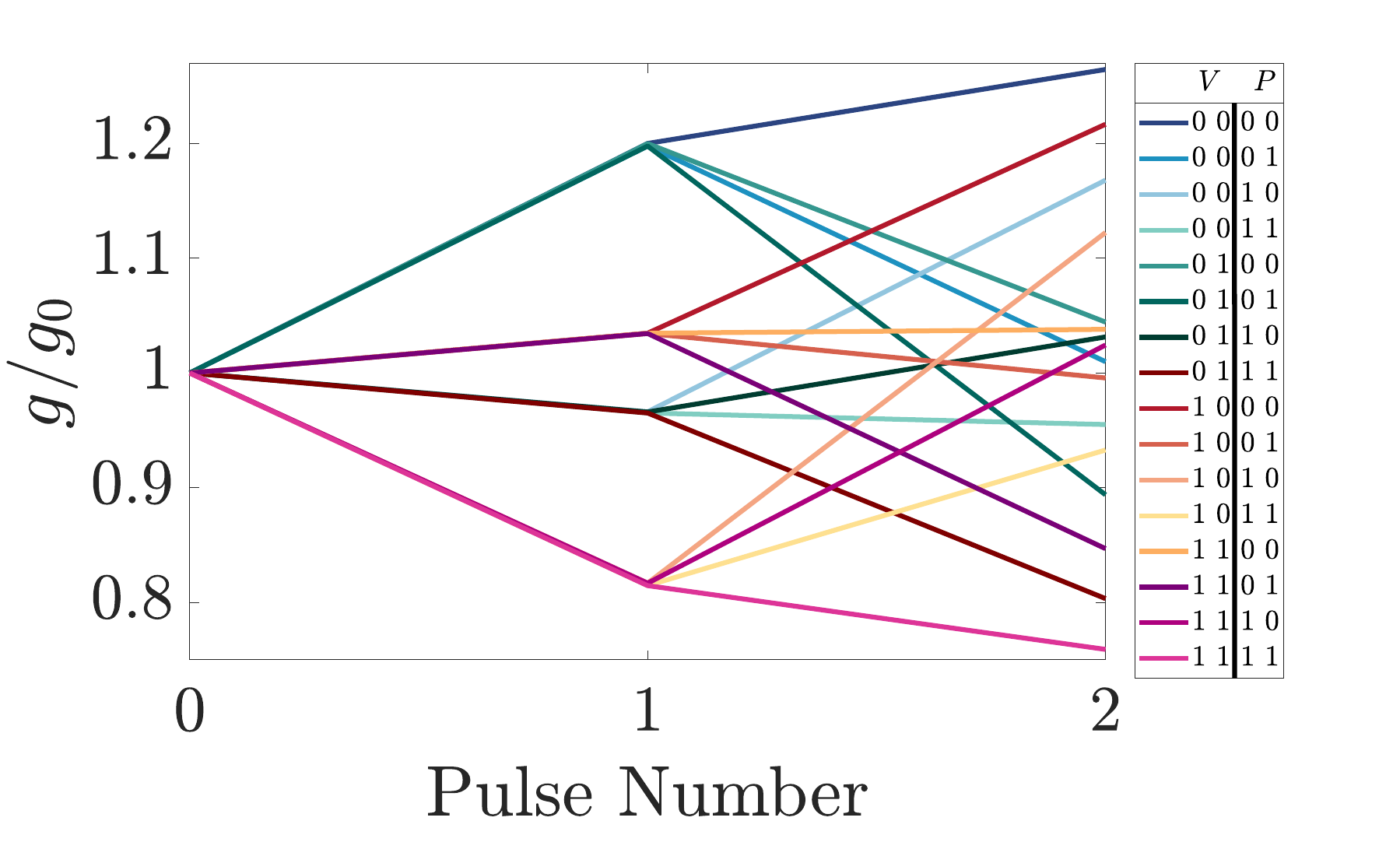}
	\caption{Relative channel conductance $g(t)/g(0)$ at $t/\tau=0.3$ and $0.6$ for all 16 possible voltage-pressure trains of two voltage-pressure block pulses leading to 16 different resulting conductances at $t=0.6\tau$. 
 } 
 \label{Fig:Conductance_2bits}
\end{figure}

Instead of a coupled voltage-pressure pulse $\pm(V_0,P_0)$, we can also consider a decoupled pulse $\pm(V_0,\pm P_0)$,  which can represent four states such that a train of only two rather than four pulses yields $4^2=16$ distinct combinations (see SM). For all 16 combinations we calculate the conductance after the first and second block pulse using the same pulse and train protocol as before, however with slightly longer pulses and separation intervals of $0.15\tau$ (rather than $\tau/8$ in Fig.~\ref{Fig:Conductance_4bits}) such that this train of two pulses only takes a time $0.6\tau$ (rather than $\tau$). The result, shown in Fig.~\ref{Fig:Conductance_2bits}, indeed reveals 16 distinct values of the conductance $g(0.6\tau)$ after the second pulse, with a window and a degree of separation that is very similar to that of Fig.~\ref{Fig:Conductance_4bits}. 
The simultaneous transmission and processing of an electro-acoustic signal increases the channel bandwidth, consequently shortening the signaling duration from 4 to only 2 pulses to reach 16 states.
Additionally, combining voltage- and pressure pulses also benefits from the enhanced range of attainable conductances, as we saw in Fig.~\ref{Fig:Conductance_4bits}, thereby increasing the capacity for distinguishing different conductance states, resulting in a multiplicative effect for information transfer rate through a conical channel. Our findings on opportunities of manipulation and signalling with pressure pulses should apply to any aqueous electrolyte in a conical channel and may represent a significant step to obviate the need for dedicated pressure sensors \cite{liu2023near}.

In conclusion, we theoretically predict on the basis of conventional transport theory for aqueous electrolytes in a conical microfluidic channel, that pulsatile flow driven by a time-dependent pressure drop can be exploited to modify and enhance the time series processing of iontronic memristors. We show how pressure pulses can either enhance the distinction of voltage signals or be used as an independent second input to achieve simultaneous electrical and mechanical time series processing, consequently doubling the channel information processing bandwidth. The classification of such time series was already shown to be applicable for iontronic fluidic reservoir computing \cite{kamsma2023brain} and therefore our results here offer a direct relevance to future development of iontronic computing. No specific features are taken into consideration apart from the standard properties of conical channels, rather our results are an intrinsic feature that should materialise in any of the widely available experimental realisations of conical channels \cite{cheng2007rectified, siwy2006ion, bush2020chemical, jubin2018dramatic, siwy2002rectification,fulinski2005transport,siwy2005asymmetric,kovarik2009effect, lin2018voltage, siwy2003preparation, siwy2002fabrication}. Consequently we provide a clear experimentally accessible method of exploiting the unique feature of a mobile medium (water) into tangible information processing benefits.

\bibliographystyle{apsrev4-2}
\bibliography{MS}

\appendix
\renewcommand*\appendixpagename{Appendix}
\appendixpage

\setcounter{figure}{0}
\renewcommand{\figurename}{Figure}
\renewcommand{\thefigure}{A\arabic{figure}}
\renewcommand{\theequation}{A\arabic{equation}}

\section{Divergence-free time-dependent currents and flow}
\begin{figure} [ht!]
\captionsetup[subfigure]{labelformat=empty}
\subfloat[\label{Fig:Q_x_Dependence}]{\subfigimg[width=\linewidth,pos=tl,vsep=1.30cm,hsep=1.25cm]{\large (a)}{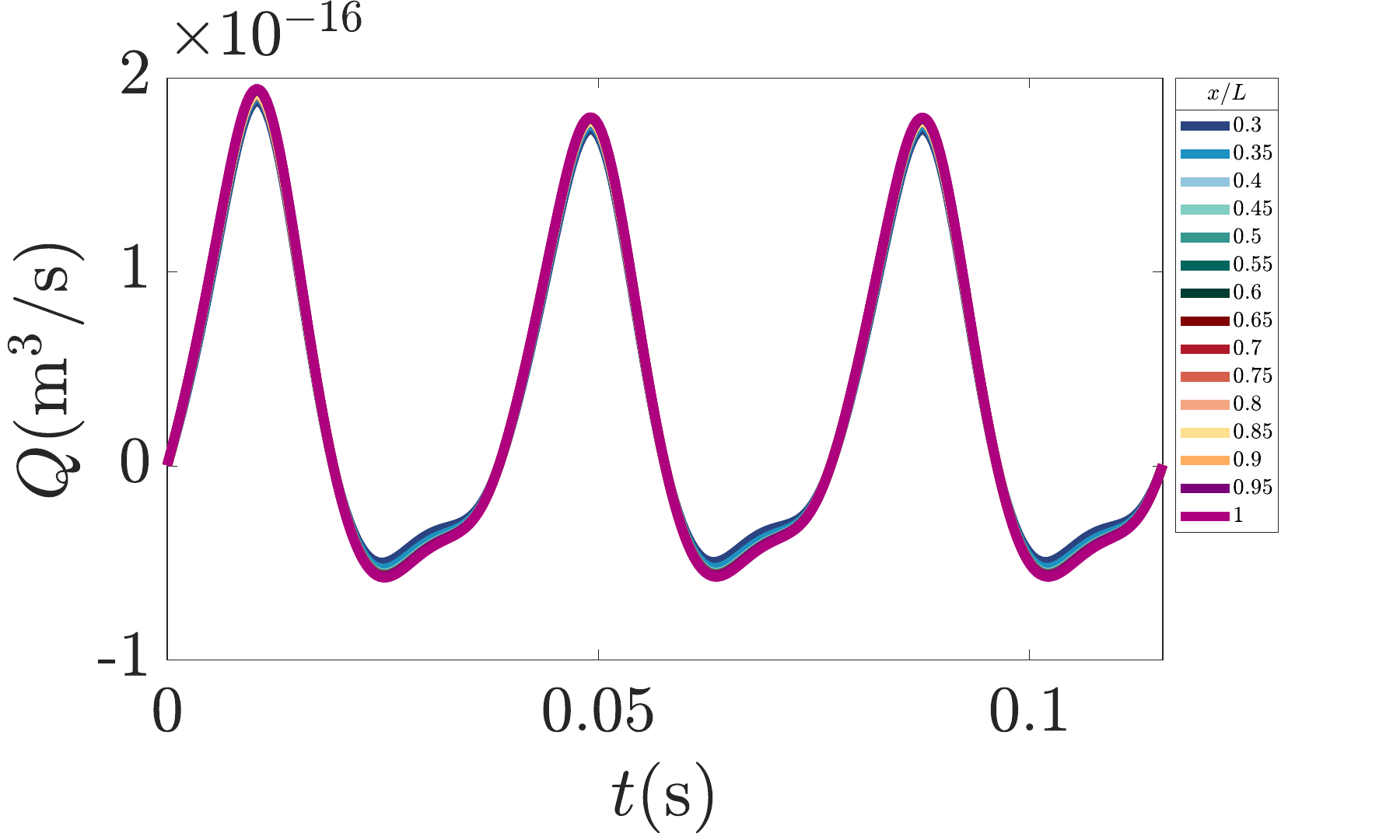}}\vspace{-0.87cm}\quad
\subfloat[\label{Fig:I_x_Dependence}]{\subfigimg[width=\linewidth,pos=tl,vsep=1.30cm,hsep=1.55cm]{\large (b)}{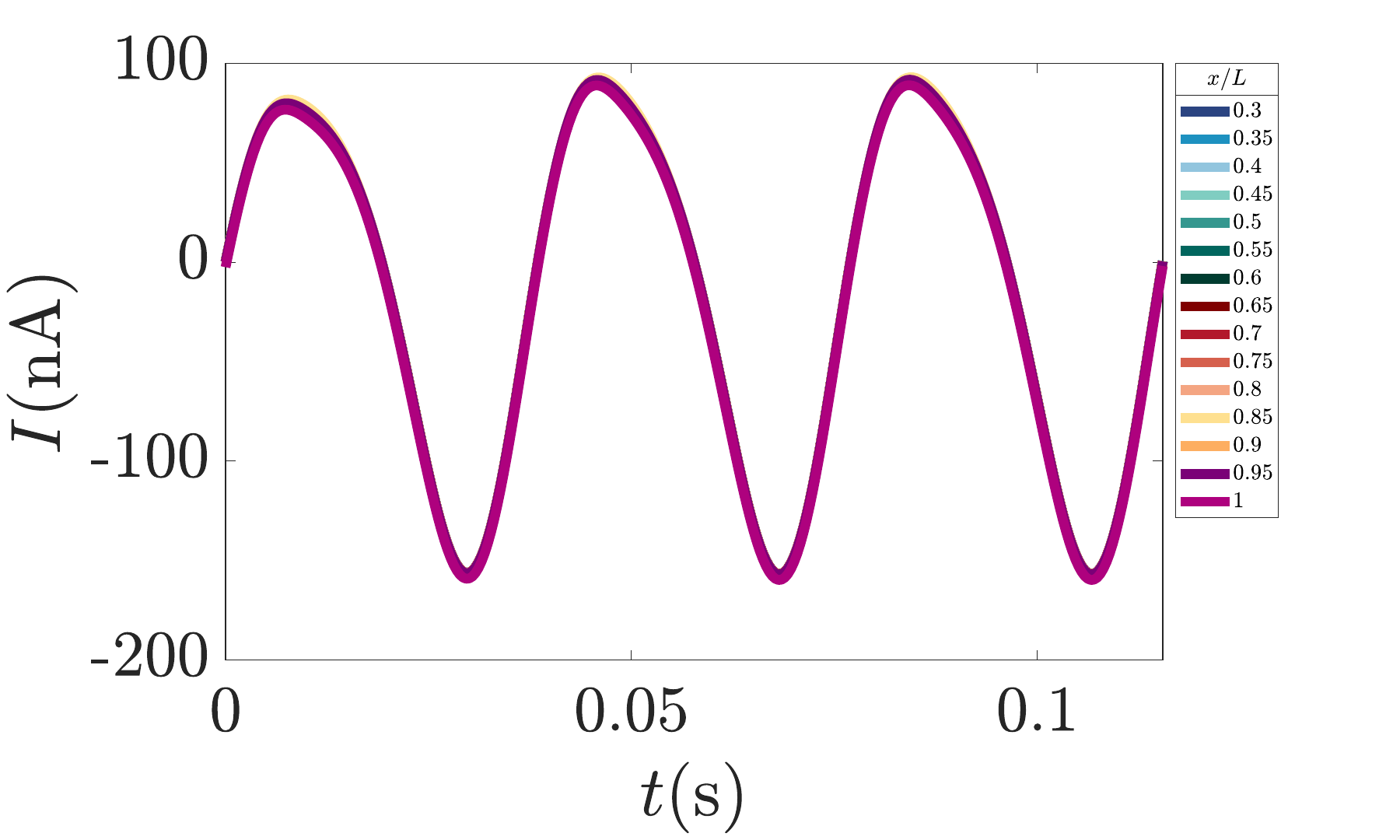}}
\caption{ Time dependence of (a) the volumetric flow $Q(t)$ and (b) the electric current $I(t)$ during the limit cycle of a periodically driven conical channel as obtained at different positions $x\in[0,L]$ in the channel for our standard parameter set (see text). The flow and current are driven by a simultaneously applied drop of the voltage $V(t)=V_0\sin(\omega t)$ and pressure  $P(t)=P_0\sin(\omega t)$ with $V_0=1~\text{V}$, $P_0=-60~\text{mbar}$ and driving frequency $\omega/2\pi=25~\text{Hz}$.  \label{Fig:QI_x_Dependence}}
\end{figure}

The electric current $I(t)$ and the volumetric fluid flow $Q(t)$ through the conical channel described in the main text and schematically shown in Fig.1 are defined in terms of the radially integrated $x$-components of the ionic fluxes ${\bf J}_\pm(x,r,t)$ and the fluid velocity ${\bf u}(x,r,t)$, such that
\begin{eqnarray} 
\begin{pmatrix} Q(t) \\ I(t)\end{pmatrix}=\int_0^{R(x)} \begin{pmatrix}{\bf u}(x,r,t)\cdot\hat{\bf{x}} \\ e\big({\bf J}_+(x,r,t) - {\bf J_-}(x,r,t)\big)\cdot\hat{\bf{x}}\end{pmatrix} 2\pi r \dd r.
\end{eqnarray}
Strictly speaking these expressions depend on the position in the channel $x\in[0,L]$, where we recall that the radius of channel is given by $R(x)$ that decreases linearly from a base radius $R_b$ at $x=0$ to a tip radius $R_t$ at $x=L$ with $L$ the length of the channel (see also main text). The incompressibility condition of the flow, $\nabla\cdot{\bf u}=0$ as imposed in Eq.(4), combined with the no-slip and no-flux boundary conditions guarantee that $Q(t)$ is independent of $x$. To a high degree of accuracy this is confirmed by our solutions of the PNPS equations, that we present in the main manuscript and here for our standard parameter set given by $L=9.8~\text{$\mu$m}$, $R_b=446~\text{nm}$,  $R_t=98~\text{nm}$ and an aqueous electrolyte with Debye length $\lambda_D=9.8~\text{nm}$ and ionic diffusion coefficient $D=1~\text{$\mu$m$^2$/ms}$ such that memristor retention time equals $\tau=8.33~\text{ms}$ according to Eq.(5). For periodic driving by a simultaneous pressure $P(t)=P_0\sin(\omega t)$ and voltage $V(t)=V_0\sin(\omega t)$, with $V_0=1~\text{V}$, $P_0=-60~\text{mbar}$, and frequency $\omega/2\pi=25~\text{Hz}$, we show the resulting (limit cycle of) $Q(t)$ at many positions between $x/L\in[0.3,1]$ in Fig.~\ref{Fig:Q_x_Dependence}. The collapse of these curves, which is excellent for $0.5<x/L<1$ for all times, is the hallmark for incompressibility of the flow, and the small deviations from collapse in the wider part of the channel at $x/L<0.5$ can be attributed to the finite grid of the numerical calculations. For the same system parameters and driving, we plot the electric current $I(t)$ for the same set of positions in the channel in Fig.~\ref{Fig:I_x_Dependence}. Interestingly, whereas some degree of $x$-dependence of $I(t)$ could in principle be possible, the essentially perfect collapse of all curves indicates that the (differential) capacity of the channel is apparently so small that no sign of a net charge build up can be detected anywhere in the channel. We conclude, therefore, that we can speak of \textit{the} fluid flow $Q(t)$ and \textit{the} current $I(t)$ through the channel, rather than of their local analogues that depend (weakly if at all) on the position in the channel.

\begin{figure} [ht!]
\captionsetup[subfigure]{labelformat=empty}
\subfloat[\label{Fig:4Bit_Pulse}]{\subfigimg[width=\linewidth,pos=tl,vsep=0.95cm,hsep=1.45cm]{\large (a)}{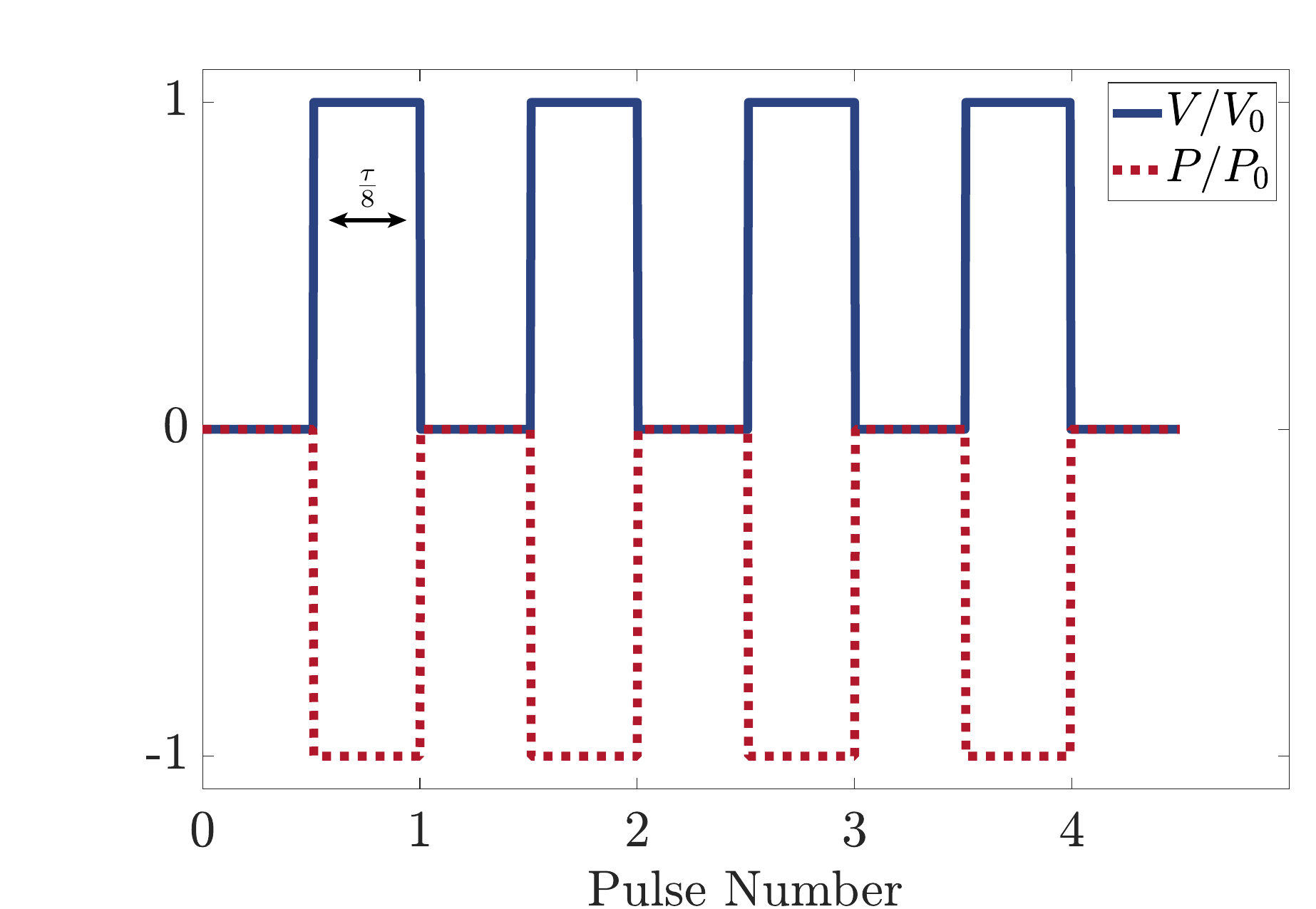}}\vspace{-0.87cm}\quad
\subfloat[\label{Fig:Conductance_Measurement}]{\subfigimg[width=\linewidth,pos=tl,vsep=0.75cm,hsep=1.45cm]{\large (b)}{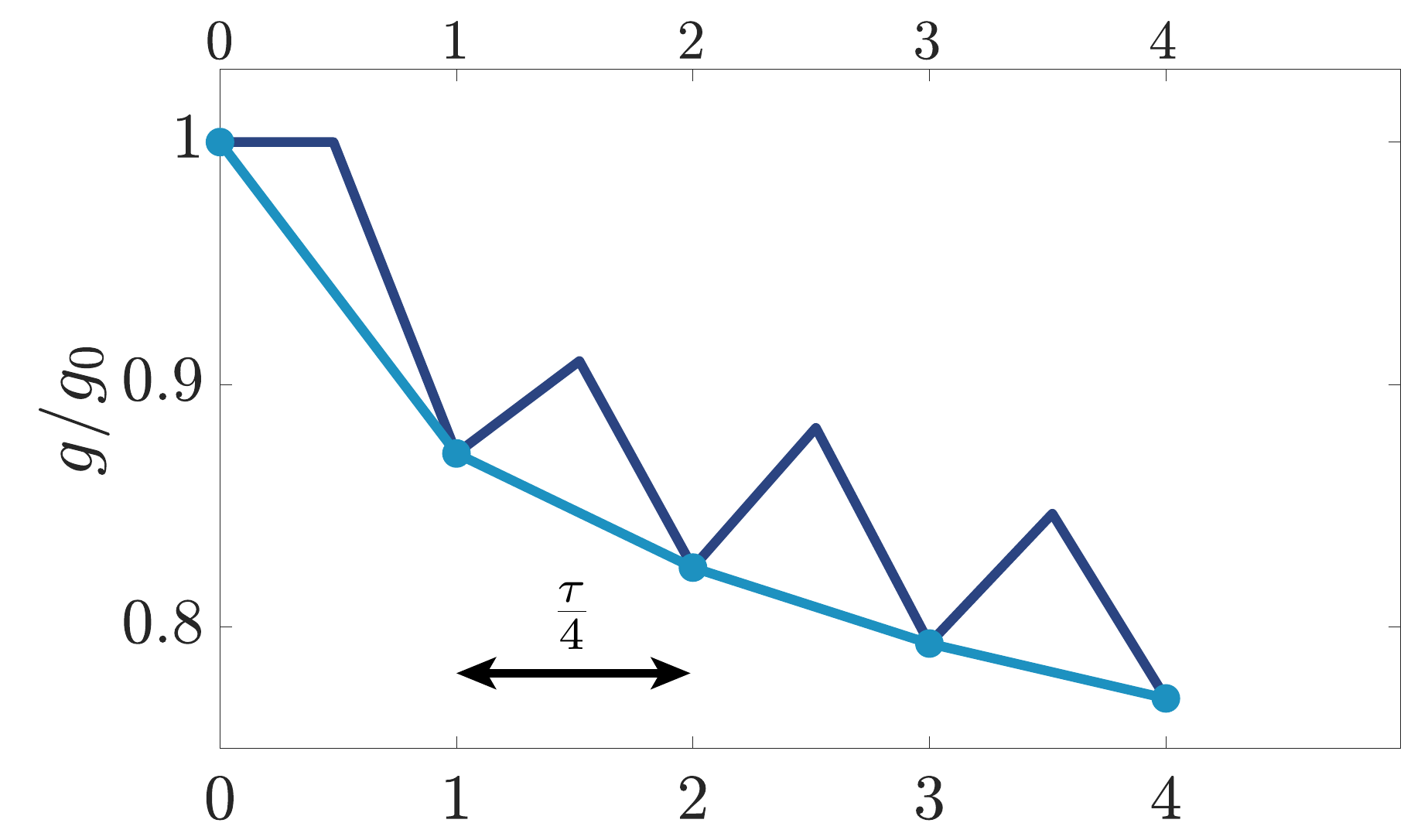}}
\caption{ (a) Example of an external driving signal, representing the bit-string 1111 and consisting of a train of $4$ block pulses $+(V_0,P_0)$ of positive voltage $V_0$ and negative pressure $P_0$, each of duration $\tau/8$ and separated by an interval $\tau/8$ with $\tau$ the typical retention time of salt in the channel. The magnitude of voltage and pressure ensure ideal conditions for salt depletion, which implies for our standard parameter set and the present case of $V_0=1~\text{V}$ that $P_0=-60~\text{mbar}$. (b) Evolution of the normalized channel conductance $g(t)/g_0$ (dark blue) under the influence of the driving of (a) that lowers the salt concentration in the channel, with $g_0=g(0)$ the initial equilibrium conductance at zero voltage and zero pressure. The connected light blue dots represent the (scaled) conductance at the very end of each block pulse, which are time-separated by $\tau/4$.  \label{Fig:Pulses_4bits}}
\end{figure}

\section{Pressure-induced widening of the conductance window}
In Fig.~\ref{Fig:Conductance} of the main text we consider the time-dependent channel conductance $g(t)$ due to simultaneously applied sequences of voltage and pressure block pulses, each of duration $\tau/8$ and magnitude $\pm(V_0,P_0)$ with $V_0=1~\text{V}$ and either $P_0=-60~\text{mbar}$ or $P_0=0$. Here a single block pulse $-(V_0,P_0)$ can represent a binary "0" and $+(V_0,P_0)$ a binary "1". If each of these pulses is preceded by a period $\tau/8$ with zero voltage and pressure, a train of four pulses takes a time $\tau$ and can represent any of the 16 binary numbers between 0000 and 1111, where the latter is represented for the case $P_0=-60~\text{mbar}$ in Fig.~\ref{Fig:4Bit_Pulse}. For our standard parameter set, and for $V(t)$ and $P(t)$ representing the ``1111'' signal of Fig.~\ref{Fig:4Bit_Pulse}, we solve the PNPS equations and calculate the time-dependent current $I(t)$, from which the time-dependent channel conductance $g(t)=I(t)/V(t)$ follows. Here we use small and short voltage ``read'' pulses in between the block ``write'' pulses, such that $V(t)$ and $I(t)$ are not strictly zero and $g(t)$ is also (numerically) well defined during several short time intervals in between the block pulses.  In Fig.~\ref{Fig:Conductance_Measurement} we plot (dark blue line) the relative channel conductance $g(t)/g(0)$ associated with the 1111 signal of Fig.~\ref{Fig:4Bit_Pulse}. We see a progressive reduction of $g(t)$ during the block pulses and a tendency to relax back towards $g(0)$ in between the block pulses. This effect is attributed to ionic depletion and re-accumulation in the channel upon applying positive and zero voltages, respectively, occurring on the time scale $\tau$ so only taking place partially during the pulses and their intervals of duration $\tau/8$. The five connected light blue dots in Fig.~\ref{Fig:4Bit_Pulse} represent $g(t)/g(0)$ at $t=0$ and at the very end of the $n$-th pulse for $n=1,2,3,4$ at times $t=n\tau/4$, showing a steady decrease with $n$ towards $g(\tau)/g(0)\simeq 0.8$.

\begin{figure}[h] 
	\centering
	\includegraphics[width=0.99\linewidth]{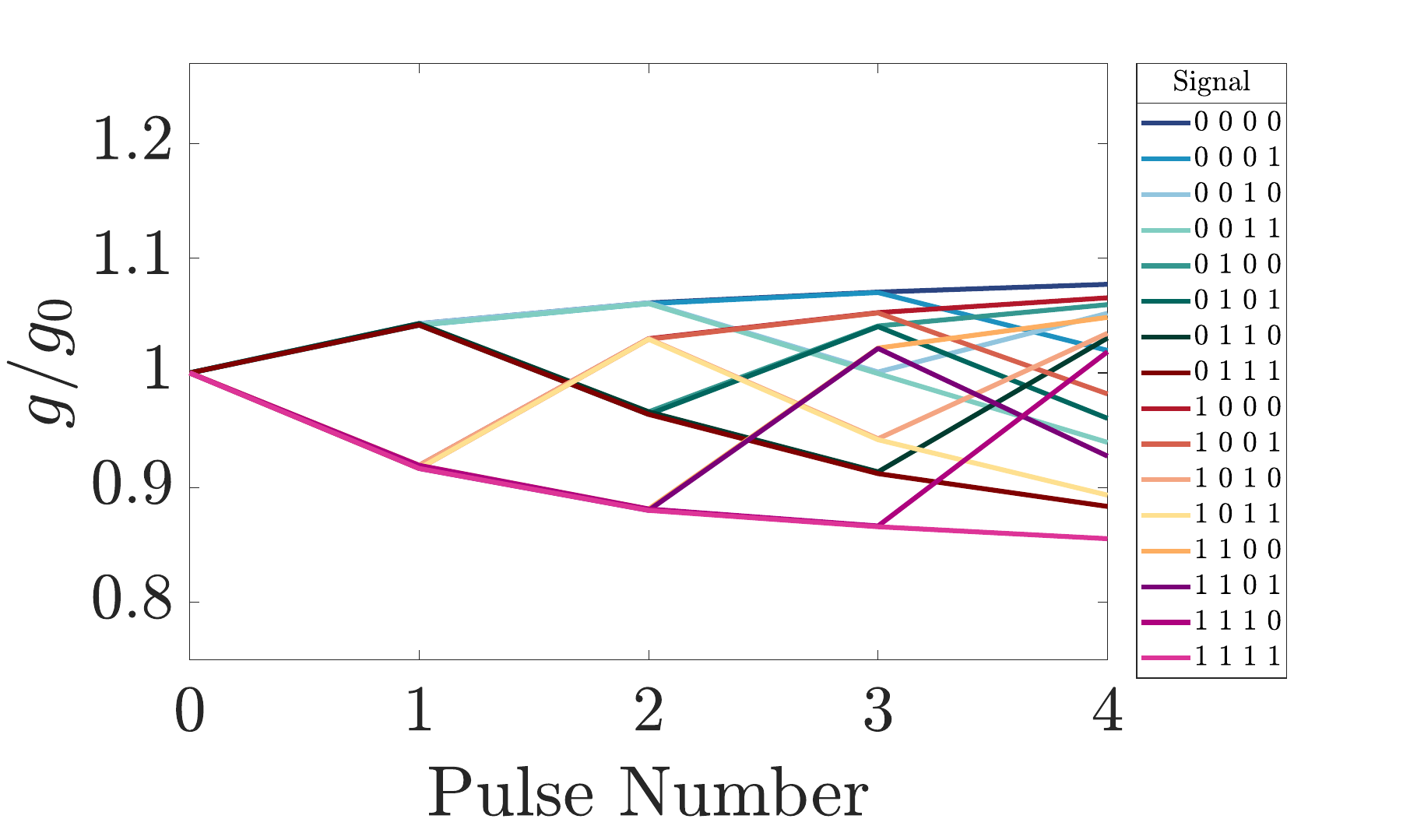}
	\caption{Relative channel conductance $g(t)/g_0$ at $4t/\tau=0,1,2,3,4$ for all 16 combinations of trains (of duration $\tau$) of four voltage pulses ($V_0=\pm 1~\text{V}$, duration $\tau/8$) without pressure ($P_0=0$). The scale of the plot is identical to the scale of Fig.~\ref{Fig:Conductance_4bits} of the main text to emphasize the relatively narrow conductance window in the absence of pressure pulses.} \label{Fig:A3}
\end{figure}

In Fig.~\ref{Fig:Conductance_4bits} of the main text we present the channel conductance $g(t)$ for $4t/\tau=1,2,3,4$, so  directly after each of the four pulses of the train, not only for the 1111 signal shown in Fig.~\ref{Fig:4Bit_Pulse} but for all 16 possible trains at $P_0=-60~\text{mbar}$. In Fig.~\ref{Fig:Conductance_Measurement} we present the same plot for the case with voltage pulses without simultaneous pressure pulses, so for $P_0=0$. This explicitly shows that the window of $g(t)$ with voltage-only pulses is substantially narrower than with pressure pulses of magnitude $P_0=-60~\text{mbar}$, while the ordering of $g(t)$ at zero and at finite pressure remains the same. One checks by a comparison with Fig.~\ref{Fig:Conductance_Spread} of the main text that the window of the conductance is set by the two extreme pulses representing 0000 and 1111, both for zero and for the nonzero pressure. Thus, without pressure the resulting 16 conductances at the end of the train are packed in a significantly narrower range, making it harder to resolve them. 

\section{Independent voltage-pressure pulses}
\begin{figure} 
	\centering
	\includegraphics[width=0.99\linewidth]{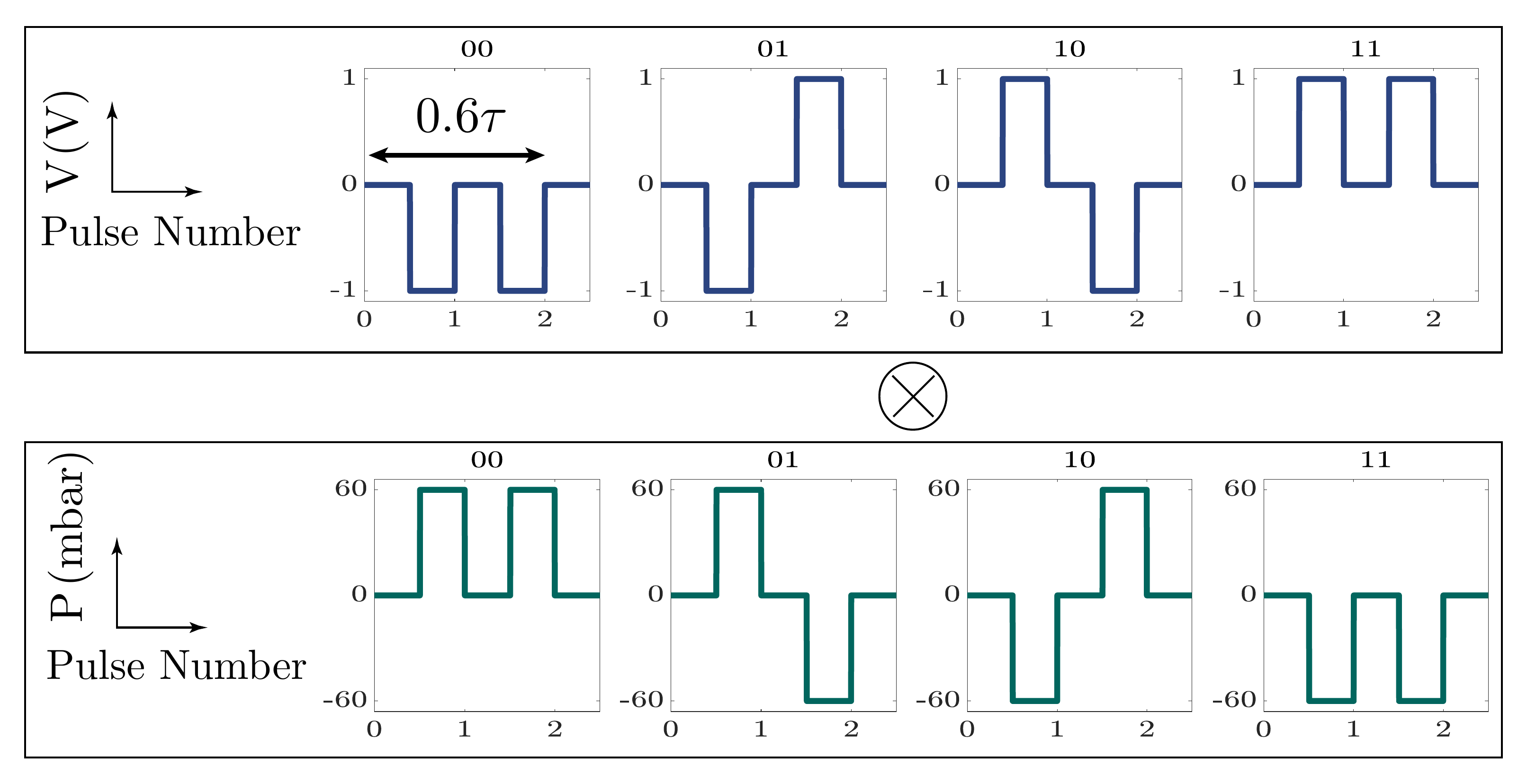}
	\caption{Representation of all $4\times4=16$ possible combinations of independent 2-pulse trains of voltage (top row) and pressure (bottom row).  The total duration of each of the simultaneously running trains is $0.6\tau$, where the duration of each pulse equals $0.15\tau$ and where each pulse is preceded by a zero voltage and zero pressure period of $0.15\tau$. } 
 \label{Fig:2Bit_Plot_SM}
\end{figure}

In the top row of Fig.~\ref{Fig:2Bit_Plot_SM} we plot the four possible combinations of 2-pulse voltage trains (representing 00, 01, 10, and 11 within our convention that a negative/positive voltage pulse represents a binary 0/1), while the bottom row represents the four possible combinations of 2-pulse pressure trains (also representing 00, 01, 10, and 11 within our convention that a positive/negative pressure represents a binary 0/1). In the case that the voltage and pressure pulses are independent, a simultaneously applied 2-pulse voltage-pressure train can thus represent $4^2=16$ combinations. The channel conductance upon applying these 16 combinations, with voltages $V_0=\pm 1~ \text{V}$, pressures $P_0=\pm 60~\text{mbar}$, and of total duration $0.6\tau$, is plotted in Fig.~\ref{Fig:Conductance_2bits} of the main text.

\end{document}